\documentclass[12pt,preprint]{aastex}

\def\ltsima{$\; \buildrel < \over \sim \;$}
\def\lsim{\lower.5ex\hbox{\ltsima}}
\def\gtsima{$\; \buildrel > \over \sim \;$}
\def\gsim{\lower.5ex\hbox{\gtsima}}

\begin{document}
\title
{Photometric Solutions for Detached Eclipsing Binaries: selection of ideal distance indicators in the SMC}

\author{J. S. B. Wyithe\altaffilmark{1,2}, R.~E.~Wilson\altaffilmark{3}}

\altaffiltext{1}{Princeton University Observatory, Peyton Hall, Princeton, NJ 08544, USA}

\altaffiltext{2}{School of Physics, The University of Melbourne, Parkville, Vic, 3052, Australia}

\altaffiltext{3}{Astronomy Department, University of Florida, Gainesville, FL 32611}

\begin{abstract}

Detached eclipsing binary stars provide a robust one-step distance determination to nearby galaxies. As a by-product of
 Galactic microlensing searches, catalogs of thousands of variable stars including eclipsing binaries have been produced
 by the OGLE, MACHO and EROS collaborations. We present photometric solutions for detached eclipsing binaries in the Small
 Magellanic Cloud (SMC)  discovered by the OGLE collaboration. The solutions were obtained with an automated version
 of the Wilson-Devinney program. By fitting mock catalogs of eclipsing binaries we find that the normalized stellar
 radii (particularly their sum) and the surface brightness ratio are accurately described by the fitted parameters
 and estimated standard errors, despite various systematic uncertainties. In many cases these parameters are well
 constrained. In addition we find that systems exhibiting complete eclipses can be reliably identified where the 
 fractional standard errors in the radii are small. We present two quantitatively selected sub-samples of eclipsing binaries
 that will be excellent distance indicators. These can be used both for computation of the distance to the SMC
 and to probe its structure. One particularly interesting binary has a very well determined solution, exhibits complete
 eclipses, and is comprised of well detached G-type, class $II$ giants. 

\end{abstract}


\section{Introduction}

The goal of the Hubble Space Telescope (HST) Key project on the extragalactic
distance scale is to measure the redshift - distance relation
 from Cepheid variables (e.g. Freedman et al. 2001). However, the distance modulus for the
 Large Magellanic Cloud (LMC) that normalizes the HST Cepheid distances
 is a matter of some debate. Because
the extrema of published results formally differ at 
the 3$\sigma$ level (Gibson 1999), it is clear that significant and unappreciated systematic
uncertainties exist in some or all of the methods.
It has been argued (Paczynski 1997, 2000) that with current technology, detached eclipsing
 binaries provide the most direct and accurate distance to the Magellanic Clouds (see Paczynski 1997 for a simple outline of the method, and Kruszewski \& Semeniuk 1999
 for an historical outlook). Two examples (for the LMC) are already in the literature (Guinan et al. 1998; Fitzpatrick
 et al. 2001), demonstrating the potential of the technique. Furthermore, a consistency check on the eclipsing
 binary method was presented by Semeniuk (2000), who compared detached eclipsing binary parallaxes
 by Lacy (1979) with Hipparcos parallaxes and found good agreement. The attributes of the most promising eclipsing
 binary systems for distance determination were discussed by Paczynski (1997). The two stars should be
 similar, in size and effective temperature, and have radii much smaller than the stellar separation.
 In addition, the inclination should be high to facilitate deep eclipses. Furthermore, the most reliable distances
 will come from objects that are subject to very small amounts of interstellar reddening. Systems where the stars
 are completely eclipsed are particularly important because they can provide robust measurements of the ratio of
 radii. In this paper we
 have restricted our attention to detached eclipsing binaries. However we note that 
 semi-detached and over-contact
 systems often have photometric solutions that are better constrained, and will
 also be useful for distance determination. We define over-contact to mean that both stars exceed
their critical lobes (\textit{viz.} Wilson, 2001).

 With high quality spectroscopic and photometric observations, standard eclipsing binary light curve fitting routines provide
 accurate masses, sizes and surface brightness ratios for the components of a double lined eclipsing binary.
 Indeed, eclipsing binaries have been a primary source of fundamental stellar data (e.g. Andersen 1991). The
 distance to an eclipsing binary follows from the dimensions thus determined plus the absolute surface brightnesses, which are inferred
 from the spectral types, either through model atmospheres or an empirical relation. Empirical studies of the
 surface brightness - color relation are based on the determination of stellar radii by interferometry
 (e.g. Di Benedetto 1998; van Belle et al. 1999). Eclipsing binaries with independently measured distances can
 also be used to calibrate the relation. Kruszewski \& Semeniuk (1999)  published a list of nearby EA-type
 eclipsing binaries with measured Hipparcos parallaxes for this purpose, and argued that it was the superior method. 

In recent years the massive photometric monitoring programs (primarily for microlensing) directed toward the
 Magellanic clouds have discovered large numbers  of variable stars including eclipsing binaries 
(e.g. Grison et al. 1995; Alcock et al. 1997; Udalski et al. 1998). These catalogs motivate systematic
 searches for rare, close to ideal systems. Alcock et al. (1997) presented photometric fits
 to 611 eclipsing binaries in the LMC based on the Nelson-Davis-Etzel model 
(Etzel 1981; Popper \& Etzel 1981). Here we use the catalog of eclipsing binaries produced
for the Small Magellanic Cloud (SMC) by the OGLE collaboration
 (Udalski et al. 1998, see that paper for details of the catalog selection), which has dense
 photometric I-band light curves for 1459 eclipsing binaries.
 The light curves are publically available from http://bulge.princeton.edu/$\sim$ogle/. 
We describe an automated light curve fitting procedure and carry out solutions
for detached OGLE SMC binaries. Our primary aim is to select detached
systems that are most suitable for distance determination. Semi-detached and over-contact
 eclipsing binaries in the SMC will be discussed in a subsequent paper. This work is
 a first step toward an eclipsing binary fitting pipeline with a realistic and
 physically motivated model.

Sec.~\ref{fitting_scheme} discusses our automated fitting scheme. In Sec.~\ref{mocksec} we display
 light curve fits to simulated eclipsing binary catalogs in order
 to gauge reliability. Sec.~\ref{OGLEsec} discusses two lists of the best detached candidates for distance
 determination selected from the OGLE catalog. These lists are compared with the
 subjective list compiled by Udalski et al. (1998). We also list solutions for the 57 candidate OGLE systems.

\section{The Fitting Scheme}
\label{fitting_scheme}

We use a public program (Wilson \& Devinney 1971; Wilson 1979, 1990, 1998) that, in its 
distributed form, requires user interaction at every iteration.
This lack of automation is a way
 to force users to examine the progress of a solution. However, while detailed fits to individual
 eclipsing binaries should be carried out interactively, the large
 number of systems (1459) investigated here makes continual interaction impractical.
We have therefore constructed an automated version of the Differential Corrections program. 

\subsection{the automated fitting scheme} 

The FORTRAN program {\em lc.f} produces a light curve from input parameters, while {\em dc.f}
 improves a solution and estimates parameter uncertainties using the method of differential corrections.
Our implementation of automation is via a shell script written in PERL that
 executes {\em dc}, as well as a number of auxiliary programs that create and update
 the input for {\em dc}, test for solution convergence, and create output using {\em lc}.
 While other methods might be faster, this approach allows {\em dc.f} to remain virtually
 unaltered. Future additions to {\em dc} can therefore be easily
 incorporated. The several morphological types of binary (\textit{i.e.} detached, semi-detached, etc.)
are handled by corresponding operational modes of {\em lc} and {\em dc}.
 The PERL script attempts to fit each object in mode 2, which
is for detached systems.
Initial parameter guesses are taken from a large pre-calculated library of eclipsing binary
 light curves. This is efficient, given the large number of objects. A best fit
 is first found for each library curve by adjusting the initial epoch, $T_0$. We assume the
 period determined by Udalski et al. (1998). The library curve
with the closest fit is then used as the initial guess (with
 luminosities scaled appropriately). We feel this to be a better approach than attempting
 to estimate the parameters from eclipse widths, positions, and depths, due to the relatively noisy data. 

 The method of multiple subsets (Wilson \& Biermann 1976) and the Marquardt algorithm
 (Marquardt 1963; Levenberg 1944) with $\lambda=10^{-5}$ were used to improve convergence. 
 Our implementation proceeds from the initial guess via sequential iterations, each
 consisting of an execution of {\em dc} for each of the 3 parameter subsets (parameters are
 updated after each subset run), followed by an execution
 with the whole parameter set (for the estimation of standard errors). Convergence is defined
 to have been achieved if parameter adjustments from each of the individual subset runs are smaller
 than 0.2 times the standard errors (calculated from running {\em dc} on the full
 parameter set). Two groups of subsets are defined. If the solution fails to converge 
 or moves to an unphysical region of parameter space with the first group, the iteration
 is terminated and convergence is sought from the initial guess for the second group.
 We demand that convergence be achieved only for the parameters corresponding to the
 surface potentials ($\Omega$), effective temperature ($T$), luminosity ($L$), and inclination ($i$).
 We found that eccentricity ($e$) and initial epoch were often much slower to converge by this
 criterion (though their adjustments are very small). These convergence criteria
 are relatively weak, but adequate for the present
 problem (see Sec.~\ref{mocksec}).

\subsection{parameter choices}

This section describes choices for the values of various parameters and
 quantities. The implications are discussed in Sec.~\ref{mocksec},
 where the light curve fitting algorithm is applied to a simulated catalog of eclipsing
 binaries. Star 1 is defined to be the star at superior conjunction near phase zero, with
 the phasing arranged so that it is the one with the deeper eclipse.
 Using {\em dc} in mode 2 for detached binaries we find
the surface potentials of stars 1 and 2, the orbital inclination,
 the temperature of star 1 (the temperature of star 2 is fixed), the orbital
 eccentricity, the luminosity of star 1 (the luminosity of star 2 is set by the
 other parameters), and an epoch of zero phase.

 Since we do not know the temperatures, we use {\em dc} with the black body
 approximation. We assume 10000 K for the temperature of
 star 2 (arbitrary choice) and a logarithmic limb darkening law. Explicitly (e.g. Van Hamme 1993)
 this law is written
 \begin{equation}
 D(\mu)=1-x(1-\mu)-y\mu\ln{\mu}
 \end{equation}
 where $D(\mu=\cos{\theta})$ is $I(\mu)/I(\mu=1)$
 at a given surface point in a direction $\theta$ with the normal. Intensities $I$ refer to
 averages over a pass-band. $x$ and $y$ are the linear and non-linear 
 limb-darkening co-efficients. From the colors
 reported in Udalski et al. (1998), the stars are expected to be primarily of O and
 B type. We therefore assume $x= 0.32$ and $y=0.18$ respectively for
 both stars. The values are from Van Hamme 1993 for I-band, with an
 effective temperature of 15000K and $\log g$ of 4.5 (\textit{cgs}).
 We use the simple reflection model (Wilson, 1990), with a single reflection (MREF=1, NREF=1),
 and assume the stars have no spots. We also assume synchronous
 rotation, that bolometric flux is proportional to surface gravity, and a
 bolometric albedo of 1. The scatter of observations is assumed to scale with the
 square root of the flux level (parameter NOISE=1 in {\em dc}).

 We assume that the initial argument
 of periastron, $\omega_0$, for star 1 (in radians) is constant in time, and that it
 takes a value of either 0 or $\pi$. This choice introduces systematic error into the
 solution for eccentricity and therefore into the stellar radii, and is particularly problematic
 where $\omega_0\approx\pm\pi/2$. However the recovered eccentricity (and hence the radii) will be
 reasonable approximations in most cases since if $\omega_0$
 is distributed randomly, then  for inclinations of 90 degrees, the projected position of
 periastron on the sky is distributed as $1/|sin\omega_0|$ (the results of the following section
 quantify this assertion). The mass ratio is assumed
 to be $q=m_2/m_1=1$, a choice that should be unimportant for well detached systems. There is
 an inherent ambiguity in partially eclipsing light curve fits between the radii of stars 1
 and 2. That ambiguity may extend to complete eclipses for noisy light curves.
 When computing the library of light curve guesses, we assume that the hotter
 star (star 1) is larger, a choice that effectively restricts our solutions. 
 We note that significantly evolved stars may not obey this condition. However
 the physically correct solution (which is essential for distance
 determination) will not usually be determined from photometry alone. The ambiguity must be resolved
 on a case by case basis with help from spectra. Importantly, the computation of a degenerate
 solution does not affect the selection of a particular binary as a potential distance indicator.
Tests revealed that the OGLE systems typically have
 third-light consistent with zero, and we have accordingly set third light to zero in our
 photometric solutions. Lists of fixed and adjusted parameters
 (as well as those for which convergence was sought) and control integers 
 are in Tabs.~\ref{tab00} and \ref{tab0}. Tab.~\ref{tab01} lists the two groups of subsets.

\section{Application of the Fitting Procedure to Simulated Catalogs of Eclipsing Binaries}
\label{mocksec}

\begin{figure}[htbp]
\epsscale{1.0}
\plotone{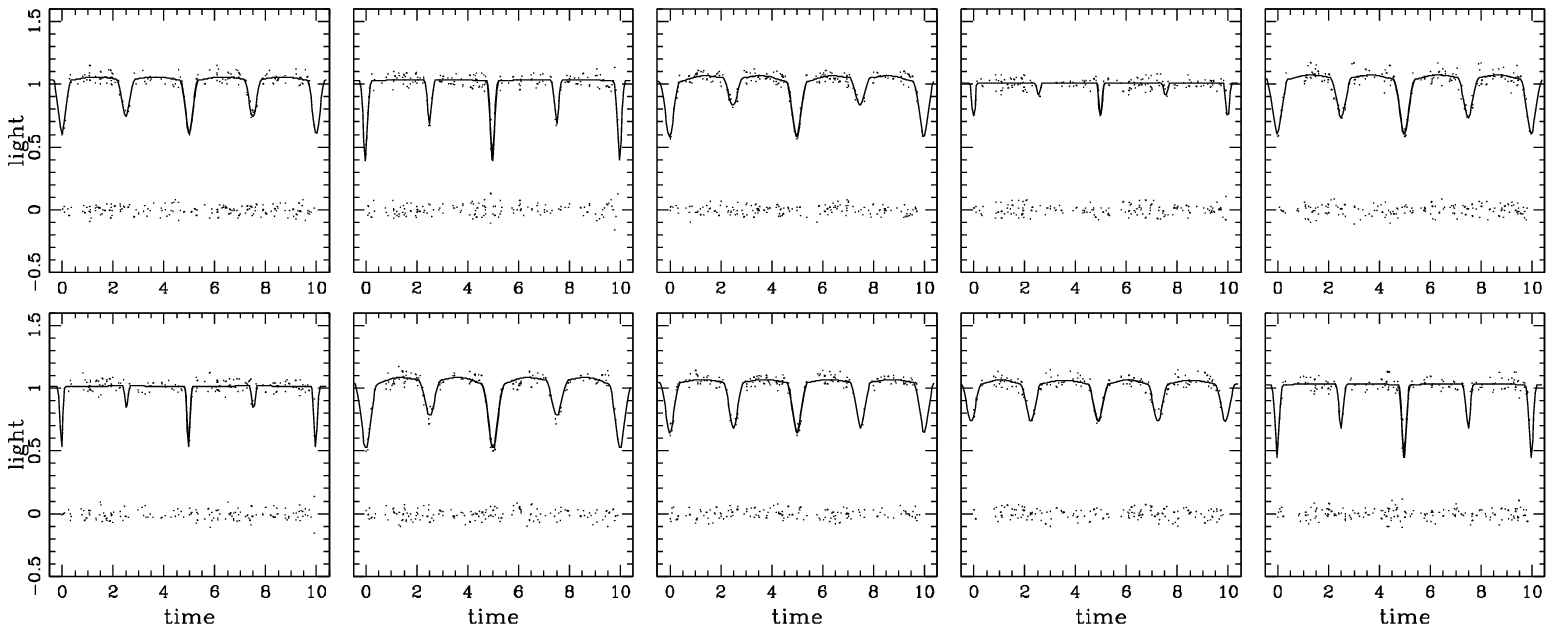}
\caption{\label{test_lc}Examples of the simulated light curve data with
 corresponding light curve solutions and residuals.}
\end{figure}

In this section we apply our light curve fitting algorithm to two simulated catalogs
 of eclipsing binaries. The first contains only detached binaries, while
 the other comprises semi-detached and over-contact systems. There are three
 objectives: 1) to quantify the applicability of the error estimates produced
 by {\em dc}, 2) to estimate the extent of the systematic error introduced by
 the assumption of fixed values for limb darkening coefficients, mass ratio,
 and argument of periastron, and 3) to estimate the success rate
 in finding good distance indicators.
 We generated 100 simulated detached eclipsing binaries having 150
 randomly spaced observations (typical for OGLE eclipsing binaries) and Gaussian
 noise that scaled as square root of light with a level of $5\%$ at the light curve mean.
These systems have temperatures for star 2 of 15000K and for star 1 of between
 15000K and 30000K. The temperature for star 2 differs from that assumed
 for real stars. We deliberately fit the wrong temperature to highlight the point that, although we do
 not know the temperatures of the real stars, the
 surface brightness ratio is the important quantity. The light curves were computed with
 logarithmic limb darkening coefficients corresponding to the temperatures (Van Hamme, 1993).
The mass ratios ranged between 0.5 and 1, and $\omega_0$
 was randomly distributed. The simulated binaries were produced
 with assorted stellar potentials, orbital eccentricities and inclinations.
 Fig.~\ref{test_lc} shows examples of the simulated light curve data, as well as
 computed light curves and residuals obtained with the fitting
 algorithm outlined in Sec.~\ref{fitting_scheme}. Our algorithm found
a converged solution with residuals comparable to the scatter
for 95\% of cases.  

\begin{figure}[htbp]
\epsscale{1.0}
\plotone{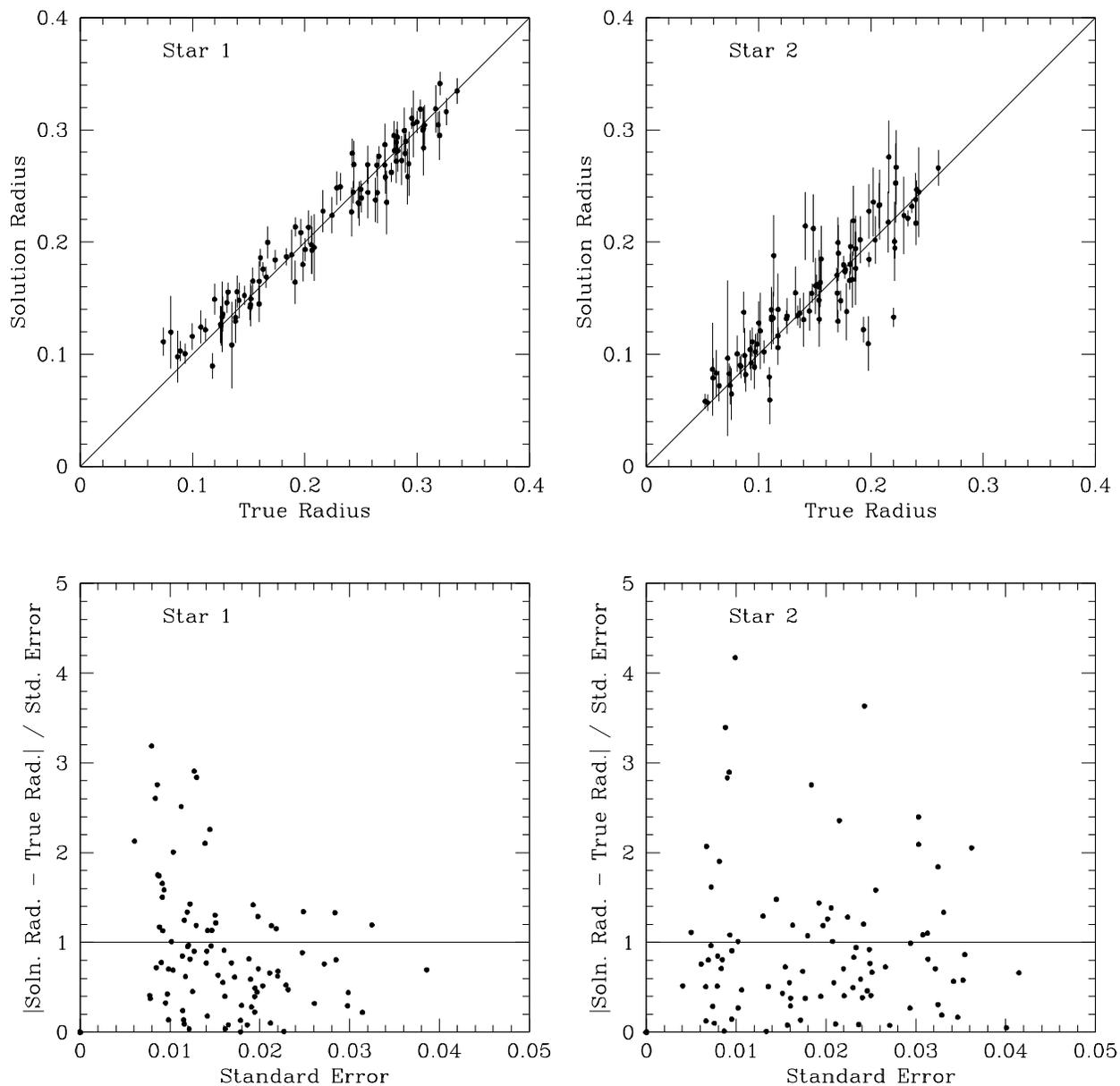}
\caption{\label{test1}Top:  Values for the radius ($r$) of Star 1 (left)
 and the radius of Star 2 (right) (with standard error, $\Delta r$) found
 from the light curve solution plotted against the values ($r_{sim})$ for
 the simulated detached eclipsing binaries. The line of equality is also
 drawn to guide the eye. Bottom: $|r-r_{sim}|/\Delta r$ plotted against the 
 standard error $\Delta r$. (Left: Star 1, Right: Star 2).}
\end{figure}

We refer to the radius toward the center of mass as the point radius ($R$).
 Radii are quoted as fractions of the stellar separation ($a$). The top
 panels of Fig.~\ref{test1} show radii of star 1 ($r_1 \equiv R_1/a$, left) and
of star 2 ($r_2 \equiv R_2/a$, right) found from the light curve solution (with
 standard error, $\Delta r$) plotted against the known
 values for the simulated binaries ($r_{sim}$). The line of
 equality is also drawn to guide the eye. The plot demonstrates that the fitting
 scheme does an excellent job of recovering input model radii, and that the
 errors provide a good description of solution accuracy (including systematic
 uncertainty). This is further demonstrated by the lower panels of Fig.~\ref{test1}
 that show $|r-r_{sim}|/\Delta r$ plotted against the absolute size of the error $\Delta r$.
 Naturally there is some tendency 
for the solutions with larger errors to be closer (in units of standard error) to the correct
 value. However the errors are surprisingly normal over the full range of error sizes,
 and do not exhibit a large tail.

\begin{figure}[htbp]
\epsscale{.75}
\plotone{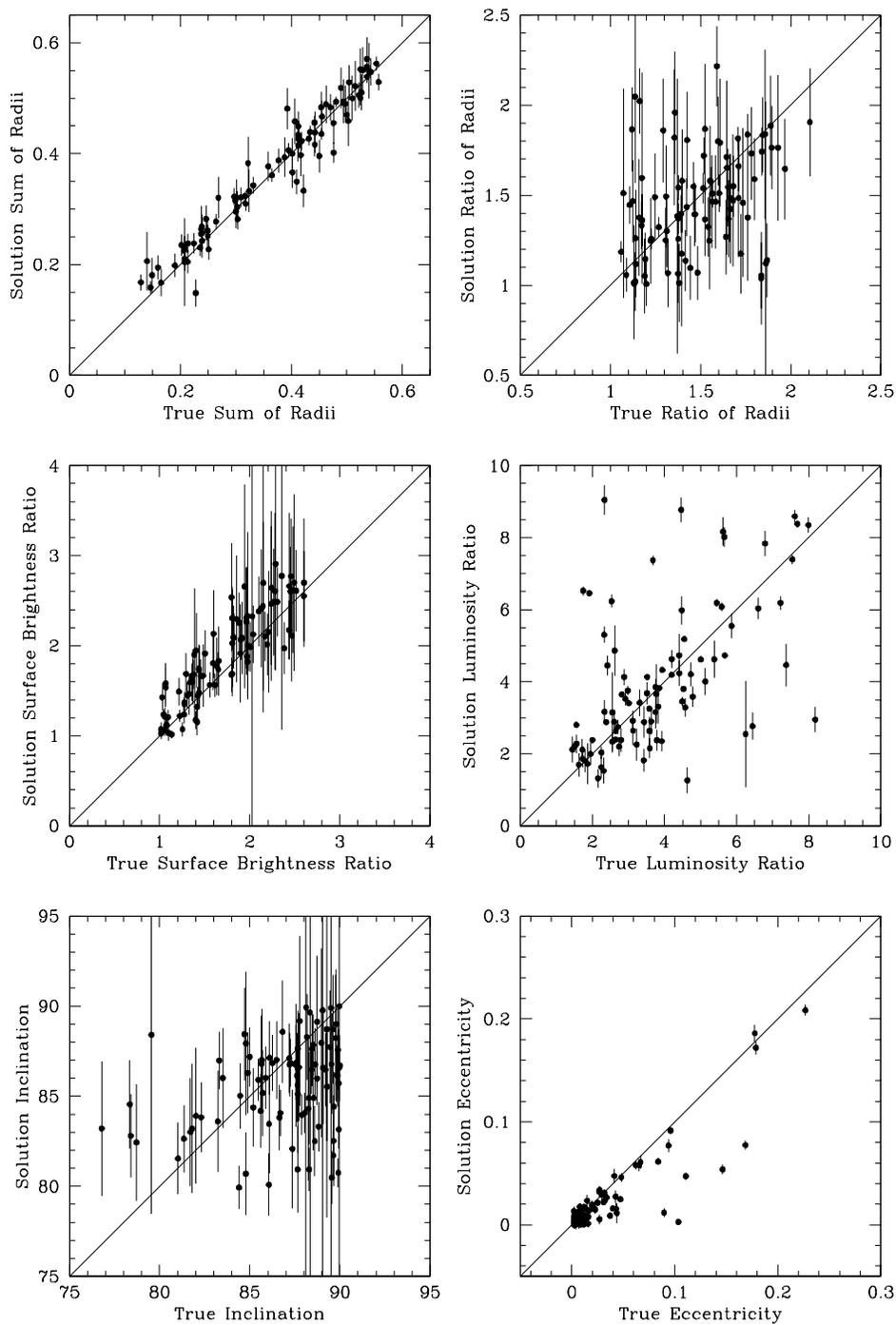}
\caption{\label{test2}Values for the light curve solution (with standard
 errors) plotted against the values for the corresponding simulated binary
 parameters. Plots are shown for the following parameters: the sum of the
 radii, the ratio of radii, the ratio of surface brightness, the luminosity
 ratio, the inclination, and the eccentricity. }
\end{figure}

Fig.~\ref{test2} shows the solution values (with standard errors) plotted against 
 simulated values for $r_1+r_2$, $r_1/r_2$,
 surface brightness ratio ($\frac{J_1}{J_2}$), luminosity ratio ($\frac{L_1}{L_2}$), inclination, and
eccentricity. The sum of radii is very accurately reproduced, while the
 surface brightness ratio is also reliably recovered. For surface
 brightness ratios near 1, the solution has a relatively small uncertainty, although
 the standard errors are realistic over the range. While the inclination is not
 recovered accurately, the standard errors are again representative. However
 significant error is clearly present in some solutions for the ratio of radii, and
 the ratio of luminosities for some systems. The eccentricity is often not recovered, and
 has unrealistic standard errors because the error is completely dominated by the
 systematic error introduced by the assumption that $\omega_0$
 is $0$ or $\pi$. The fitted eccentricity is a lower limit.

\begin{figure}[htbp]
\epsscale{1.0}
\plotone{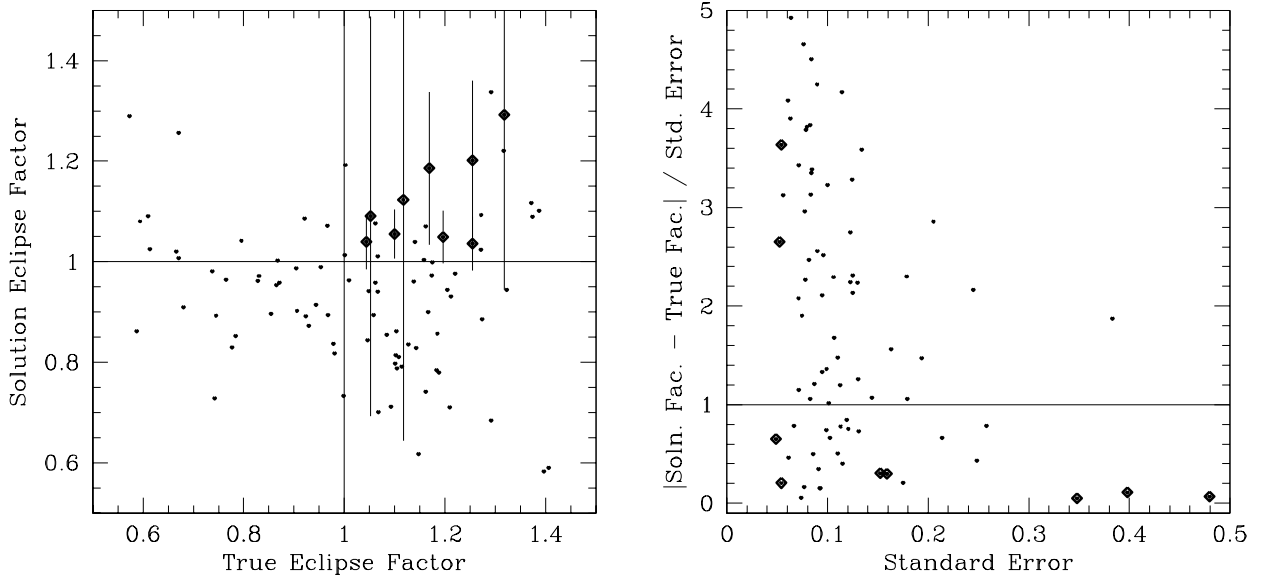}
\caption{\label{eclipse_test}Left: Values of $F_e$ found from the light
 curve solution plotted against the values for the simulated detached
 eclipsing binaries ($F_{sim}$). The lines of unity are also drawn to
 separate regions of complete and incomplete eclipse.
 Right: $|F_e-F_{sim}|/\Delta F_e$ plotted against the standard error
 ($\Delta F_e$). Cases where $\Delta r_1/r_1<0.05$ and $\Delta r_2/r_2<0.05$
 are denoted by larger dots and have their standard error bars shown
 (left panel).}
\end{figure}

 The relatively large uncertainties in the ratio of radii and inclination can
 be traced to the following: In the case of partial eclipses, a change in the ratio of radii
 can nearly be compensated (with regard to eclipse depth) by a change in inclination, so
 there is a near-degeneracy between the ratio of radii and the inclination. This degeneracy may be 
 broken if the eclipses are complete, and we thus expect smaller errors for the
 individual radii in these cases. Turning this around, solutions with small
 relative errors in
 the radii can therefore be used to determine whether a binary is completely eclipsed.
 We define the quantity:
\begin{equation}
F_e\equiv\frac{r_{p,1}+r_{p,2}-\cos{i}}{2r_{p,2}},
\end{equation}
which is greater than 1 for systems with complete eclipse (strictly true only for spherical 
 stars, but nearly true otherwise). Here the $r_p$ are the polar radii in
 units of orbital semi-major axis, $a$. 
 The left panel of Fig.~\ref{eclipse_test} shows $F_e$ for the simulated binaries plotted
 against the corresponding values from the light curve solution. Lines of unity are drawn to
 separate regions of complete and incomplete eclipse for both the simulated binaries and their solutions.
 Due to the large uncertainties in inclination and ratio of radii, the solution $F_e$'s 
 are also very uncertain. The right panel of Fig.~\ref{eclipse_test} shows the number of
 standard errors ($\Delta F_e$) by which the solution for $F_e$ differs from its correct value plotted against the
 absolute size of the error. This plot demonstrates that while $F_e$ is not recovered accurately in
 most cases, the standard errors are realistic. Cases where
 $\Delta r_1/r_1<0.05$ and $\Delta r_2/r_2<0.05$ are denoted by larger dots and have
 their standard error bars shown (left panel). The solutions for this subset of systems
 recover $F_e$ (sometimes accurately) with realistic standard errors.

 Since many of the OGLE binaries are not detached, simulated catalogs of semi-detached (mode 5) and
 over-contact (mode 1) binaries were produced by
 applying noise and sampling to model light curves as before. In the case of semi-detached binaries
 the potential of star 2 was fixed by the mass ratio. For the over-contact binaries
 the potential of star 2 was equal to the potential of star 1.
 Our detached binary fitting algorithm was unable to converge to a solution for any of the
 over-contact systems. This failure is expected and is a consequence firstly of
 the fact that the over-contact light curve geometry cannot typically be produced from a detached system, 
 and secondly that we have assumed $q=1$. Whilst the assumption of $q=1$ is reasonable when discussing
 detached systems, over-contact binaries have light curves that are sensitive to
 $q$, which is nearly always far from 1. However detached solutions were
 found for about two thirds of semi-detached binaries (the somewhat low success rate can again be
 traced to the fixed $q$). 
 Because the simulated semi-detached catalog was produced in mode 5, the cooler
 star (star 2) fills its critical lobe. The radius is therefore typically
 larger for star 2 than for star 1. However our procedure provides an initial
 guess having a radius that is larger for star 1 than for star 2.
 Solutions can be nearly ambiguous between cases where the radii but not the temperatures
 are interchanged. Since star 1 is defined to be hotter and we start from an assumption 
 that $R_1>R_2$, we may find this alternate (incorrect) solution for many SD systems when they are solved 
 as detached. Importantly, where a fit is obtained the sum of the radii and the surface
 brightness ratio are recovered accurately, and with realistic errors.
 Over-contact and semi-detached systems should therefore not contaminate
 our selection of well detached distance indicators. 

 In summary, we find the following: The assumption of an $\omega_0$
 that is either 0 or $\pi$ radians can introduce significant systematic
 error into the solution for eccentricity. The resulting systematic contributions
 to the errors in radii from the limb darkening, argument of periastron
 and the uncertainty in surface brightness ratio often render the solution for
 luminosity ratio incorrect, and with unrealistically small error bars. However
 the $r$'s, their sum, and the surface brightness ratio
 are reliably reproduced. Furthermore, solutions
 having a small relative error in both radii accurately determine $F_e$,
 and therefore whether the binary undergoes total eclipse.

\section{Application of the Fitting Procedure to the OGLE Catalog of Eclipsing
 Binaries in the SMC}
\label{OGLEsec}

We next applied our algorithm to the OGLE SMC eclipsing binary catalog.
 We obtained solutions for $\sim80\%$ of binaries. Given our high success
 rate on the simulated catalogs, we infer that the remaining $\sim20\%$
are predominantly in semi-detached or over-contact condition. Another possibility
 is that some of the un-fitted systems have mass ratios very different
 from 1, with one star having a normalized radius greater than one half.
 Not all our detached solutions necessarily imply a detached system.
 Due to the typical noise levels, a detached
 solution with $q=1$ may be found for a semi-detached light curve
 having $q\ne 1$. However, these systems will have at
 least one large component. The OGLE SMC eclipsing
 binary catalog has 68 objects that are in two overlapping fields. As a check we
found solutions for these objects using both data sets. In these
 cases we find consistent solutions.

\begin{figure}[htbp]
\epsscale{.95}
\plotone{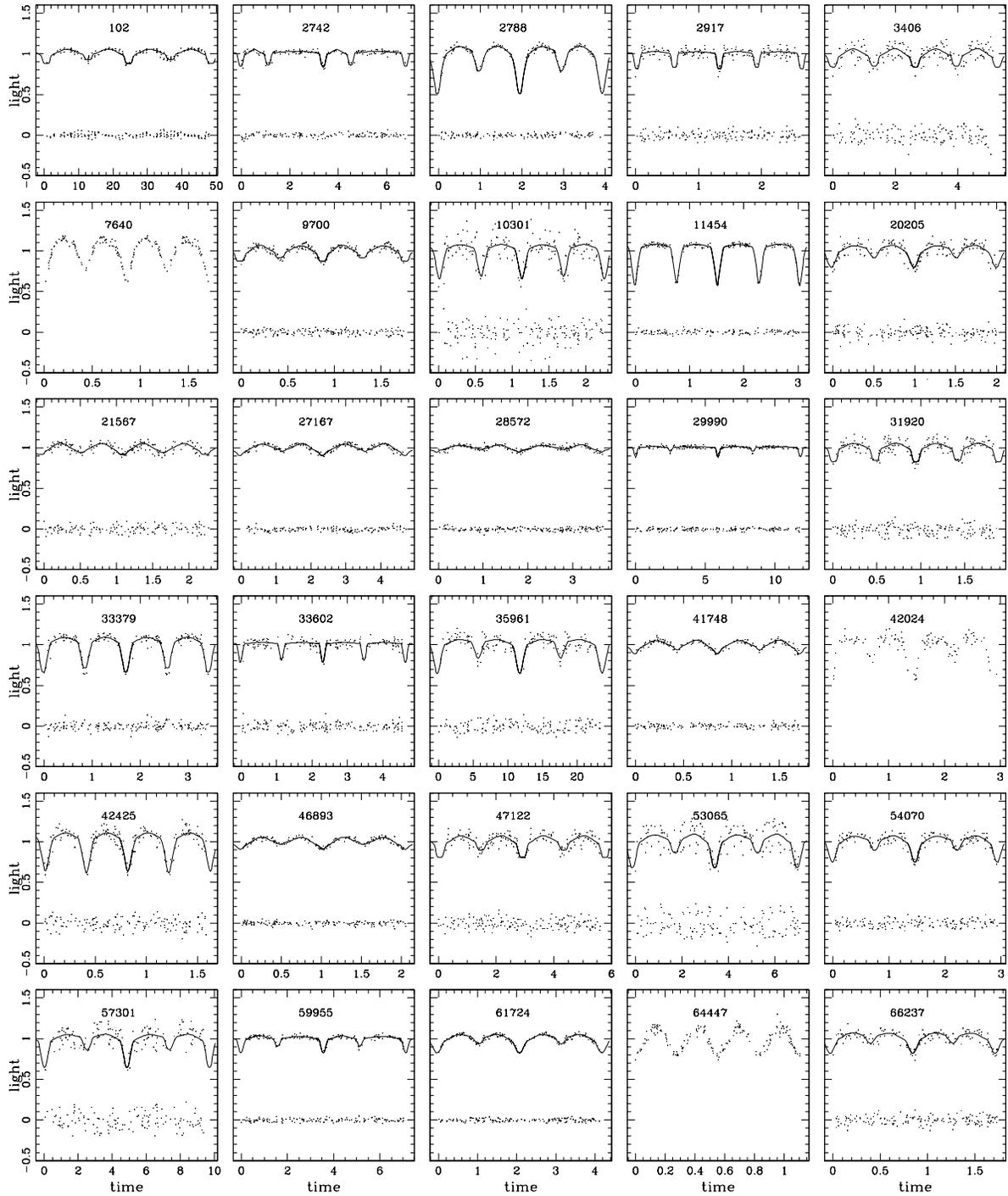}
\caption{\label{obs_fits}Examples from OGLE field smc-sc2 of light curve
 solutions with residuals, as well as systems where no solution was found
 (objects 7640, 42024 and 64447). Each panel is labeled by the OGLE object
 identification number.}
\end{figure}

\begin{figure}[htbp]
\epsscale{0.85}
\plotone{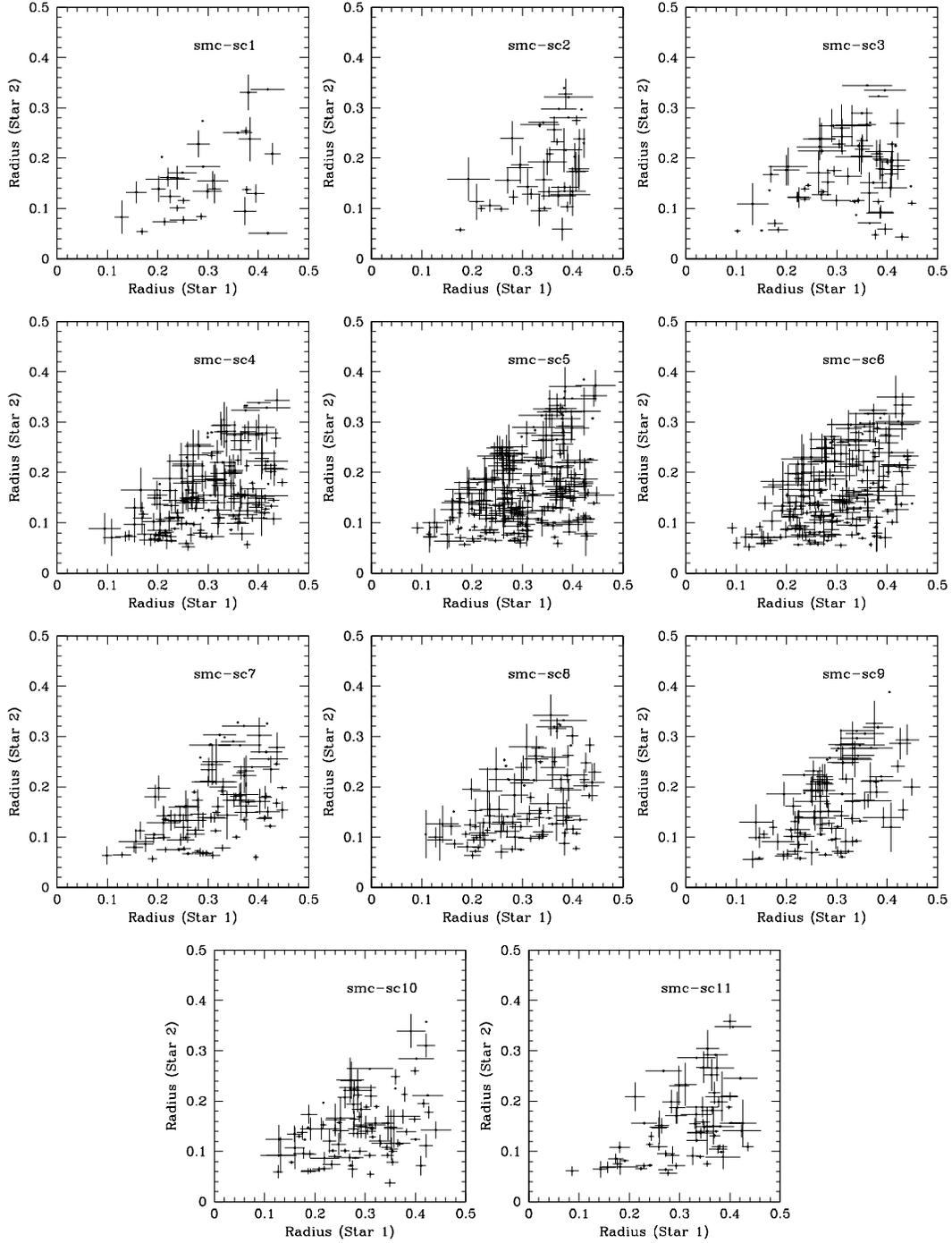}
\caption{\label{fit_param}Solutions for the radius of Star 1 plotted
 against the solutions for radius of Star 2. Only error bars smaller
 than 0.05, are shown. }
\end{figure}

\begin{figure}[htbp]
\epsscale{.85}
\plotone{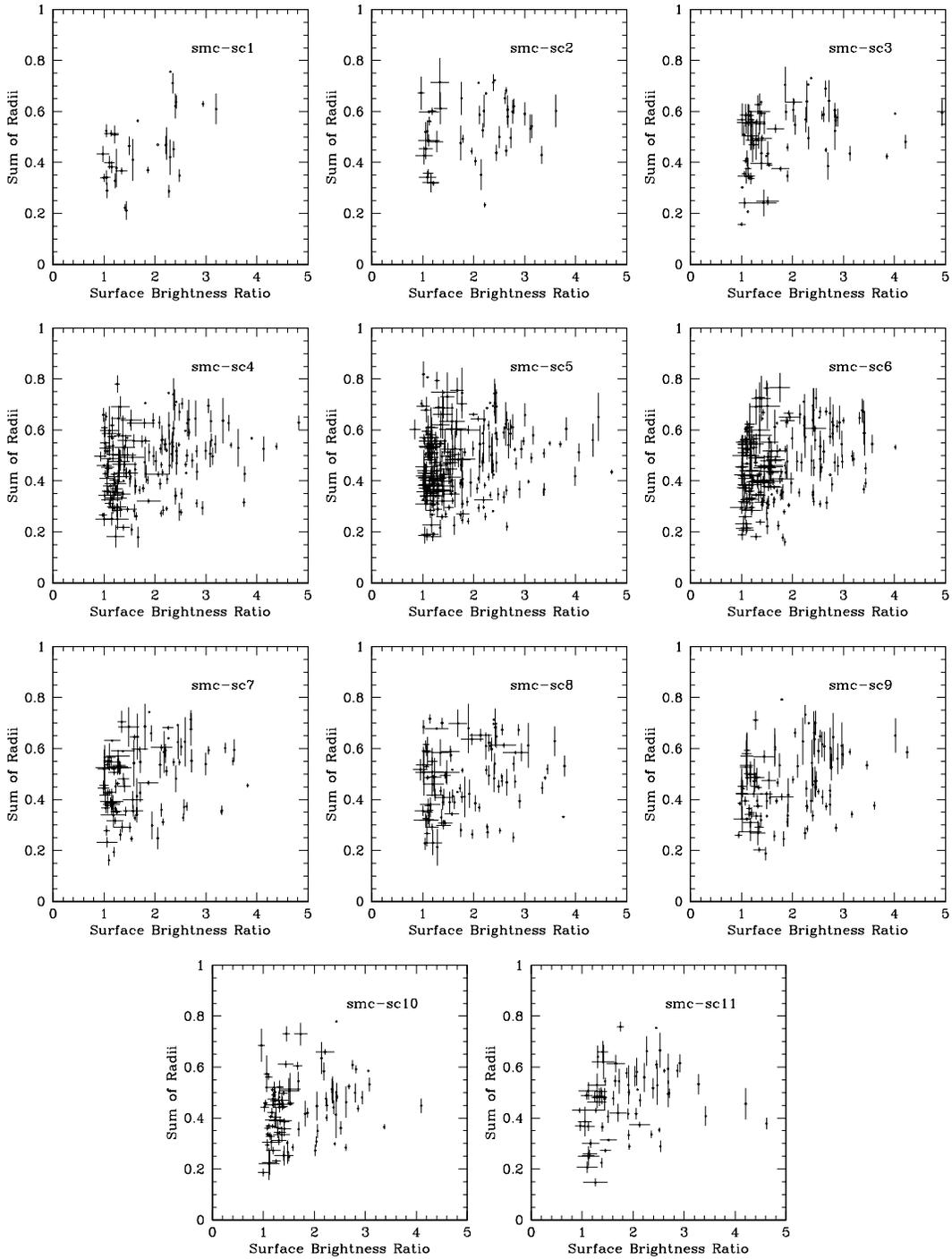}
\caption{\label{fit_param2}Solutions for the sum of the radii plotted against
 the solution for surface brightness ratio. Only error bars in surface
 brightness ratio smaller than 0.25, and in the sum of the radii smaller
 than 0.1 are shown. }
\end{figure}

Figure~\ref{obs_fits} shows 27 examples (from OGLE field smc-sc2) of
 light curve fits and residuals, as well as 3 examples 
 where no solution was found because an iteration moved outside the physical 
bounds of the $q=1$ detached system. The latter have light curves that are
strongly rounded outside the eclipses, indicative of a semi-detached
 or over-contact condition. Figs.~\ref{fit_param} and \ref{fit_param2}
 summarize the solution parameters in the OGLE SMC fields. The figures
 demonstrate the range of solutions and also the range of solution quality.
 The figures show $r_1$ vs. $r_2$ (Fig.~\ref{fit_param}) and $r_1+r_2$ vs. surface brightness
 ratio (Fig.~\ref{fit_param2}). Points for all systems are shown, but for clarity only $r_{1,2}$
 error bars smaller than 0.05,
 surface brightness ratio error bars smaller than 0.25, and $r_1+r_2$ error bars
 smaller than 0.1 are shown. Systems that will be useful distance
 indicators are principally found in the lower left regions of both plots.

\subsection{a sub-catalog of systems suitable for distance determination}

We have quantitatively selected two catalogs of objects that should be particularly reliable
 distance indicators: 

\subsubsection{well separated systems}

\begin{figure}[htbp]
\epsscale{.95}
\plotone{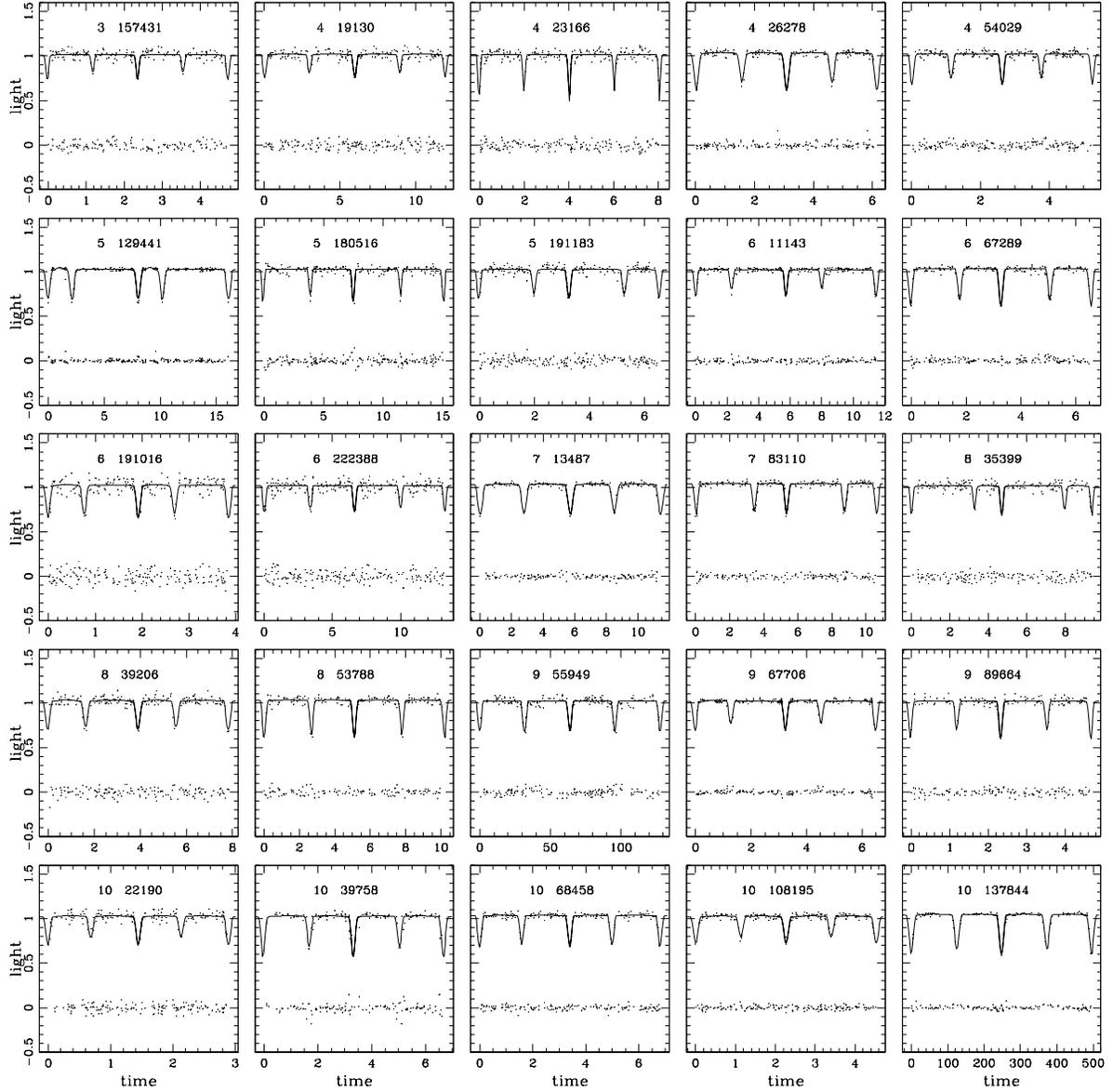}
\caption{\label{dist_indicators}A collection of light curve solutions for 
 well detached OGLE eclipsing binaries which will make the best systems
 for distance determination. Each panel is labeled by the OGLE field and
 object identification numbers. The solution parameters are given in
 Tab.~\ref{tab1}}
\end{figure}

The first catalog is composed of systems having a small and well determined $r_1 + r_2$,
 similar temperatures, and deep eclipses.
 These well detached systems have the advantage that the surface brightness - temperature
 relation can more easily be calibrated empirically. In particular, in relation to 
 surface brightness, well detached systems can be considered as two 
 isolated stars. The surface brightness - temperature relation can therefore be measured for
 isolated stars as well as eclipsing binaries. Fig.~\ref{dist_indicators} shows
 a collection of light curve fits
 for the best well detached systems
 for distance determination. Their parameters are
 in Tab.~\ref{tab1}. They have $r$'s smaller than 0.2,
 surface - brightness ratios smaller than 1.5, and sufficiently high inclinations
 so that one eclipse depth is larger than 0.25 of the maximum light curve flux
 (measured from the light curve solution). In addition the standard errors
 in radii are smaller than 0.05, the standard errors in surface brightness 
 ratio are smaller than 0.25, and the standard errors in the inclination are smaller than 10 degrees. The systems are evenly distributed across the OGLE fields. 

\begin{figure}[htbp]
\epsscale{.95}
\plotone{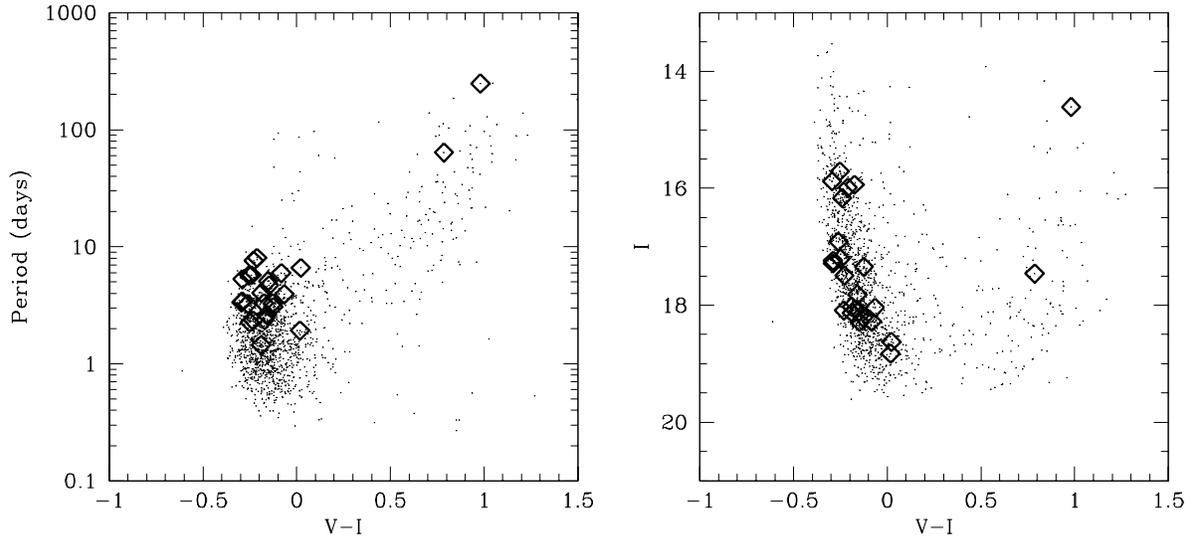}
\caption{\label{colour_mag}Color - period and color - magnitude diagrams for
 the complete eclipsing binary catalog (small dots, data from Udalski
 et al. 1998). The well detached objects selected as potential distance
 indicators are superimposed (diamonds). The V-I colors have been
 de-reddened assuming E(V-I)=0.14.}
\end{figure}

Fig.~\ref{colour_mag} shows color - period and color - magnitude diagrams
 for the selected objects superimposed
 on the full catalog of OGLE SMC eclipsing binaries.
 The V-I colors have been de-reddened assuming the average reddening of field stars for the SMC
E(V-I)=0.14 (Massey, Lang, DeGioia-Eastwood \& Garmany 1995). The potential
 distance indicators fall into two classes according to period. The great
 majority have periods of a few to 10 days, with colors and magnitudes 
consistent with their being O, B or A stars. Most of these short period
 binaries are noticeably fainter than the brightest blue binaries in the
 OGLE catalog, suggesting that they might be B or A rather than O stars.
 B-type (and cooler) components are typically more 
 desirable for distance determination than are O stars since their spectra do not usually
 contain emission lines or other peculiarities. B stars also are less likely than O stars
 to have winds that contaminate the radial velocity determination, although spectra are 
 required to establish suitability on these grounds.
 A further important consideration is the surface brightness - temperature relation,
 which one may want to calibrate empirically for the least controversial distance determination.
 The list of candidates for this calibration compiled by Kruszewski \& Semeniuk (1999)
 from the Hipparcos catalog contains no O stars. Existing calibrations 
  (e.g. Di Benedetto 1998; van Belle et al. 1999) are primarily for later spectral types.

 A second, much rarer class (objects smc-sc9 55949 and smc-sc10 137844)
 have longer periods of tens to hundreds of days, and significantly
 redder colors, consistent with G or K stars. One of these (smc-sc10
 137844) has a luminosity consistent with its components being of class II, very smooth
 residuals, and the
 smallest standard errors of any of the solutions in Tab.~\ref{tab1}. In
 addition to the original solutions, we obtained solutions for these red
 objects assuming more appropriate limb darkening coefficients ($x=0.60$, $y=0.21$,
 from Van Hamme 1993 for an effective temperature of 5000K and a
 $\log g$ of $4.5$). The results are given in parentheses in Tab.~\ref{tab1} and
 show that the two solutions are consistent to within standard error.
These are rare and potentially
 very interesting objects. They will have low radial velocity amplitudes, therefore
 requiring higher dispersion spectra. However
 they are bright, and being of a cooler spectral type, will have many spectral
 lines. Furthermore, the surface brightness - temperature relation is
 empirically well established over the range of spectral types A to
 K (Di Benedetto 1998; van Belle et al. 1999), and is nearly parallel to the reddening line
 for late type stars (Barnes \& Evans 1976). We note that none of the systems selected as
 well detached distance indicators have very short periods ($<2$ days).

\subsubsection{systems exhibiting complete eclipse} 

\begin{figure}[htbp]
\epsscale{.95}
\plotone{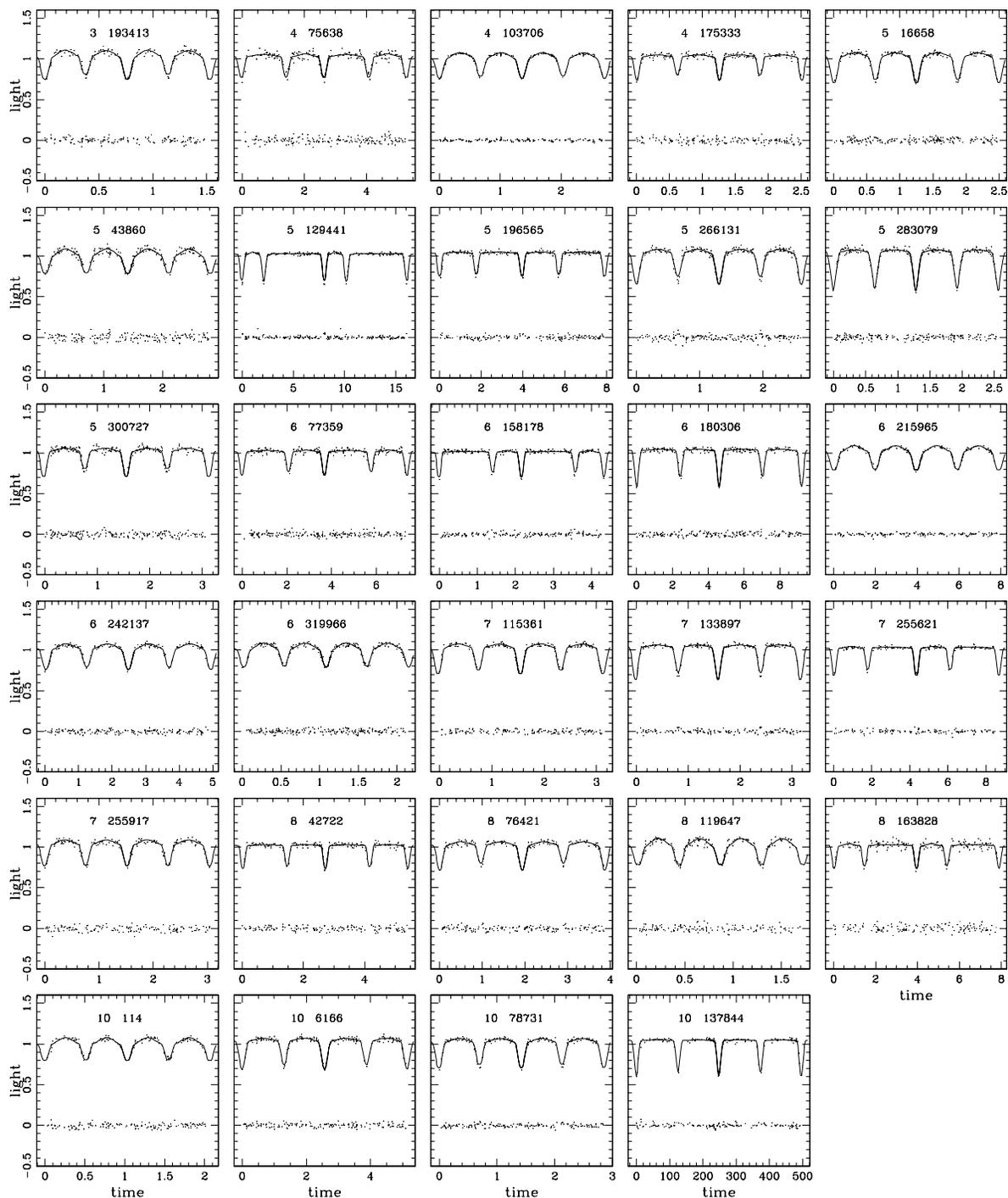}
\caption{\label{ecl_dist_indicators}A collection of light curve solutions for 
 OGLE eclipsing binaries exhibiting complete eclipse. Each panel is labeled
 by the OGLE field and object identification numbers. The solution
 parameters are given in Tab.~\ref{tab2}}
\end{figure}

 A second catalog contains detached systems with small relative errors in individual
 radii, similar temperatures, and complete eclipses. These systems have the
 advantage that the photometric solutions may often provide good determinations of inclination,
 or at least a robust lower limit.
 Fig.~\ref{ecl_dist_indicators} shows the $29$ light curve fits whose
 parameters are in Tab.~\ref{tab2}. The systems have $\Delta r/r$'s smaller than 0.05,
 surface-brightness ratios smaller than 1.5, standard errors in surface brightness
 ratio smaller than 0.25, and sufficiently high inclinations
 so that one eclipse depth is larger than 0.25 of the maximum light curve flux
 (measured from the light curve solution). Instances of complete eclipse were determined from $F_e$.
 The solutions yield either $F_e>1$ and $\Delta F_e<0.05$ or $F_e>1.1$ and $\Delta F_e<0.1$. These systems are
 again evenly distributed across the OGLE fields. 

\begin{figure}[htbp]
\epsscale{.95}
\plotone{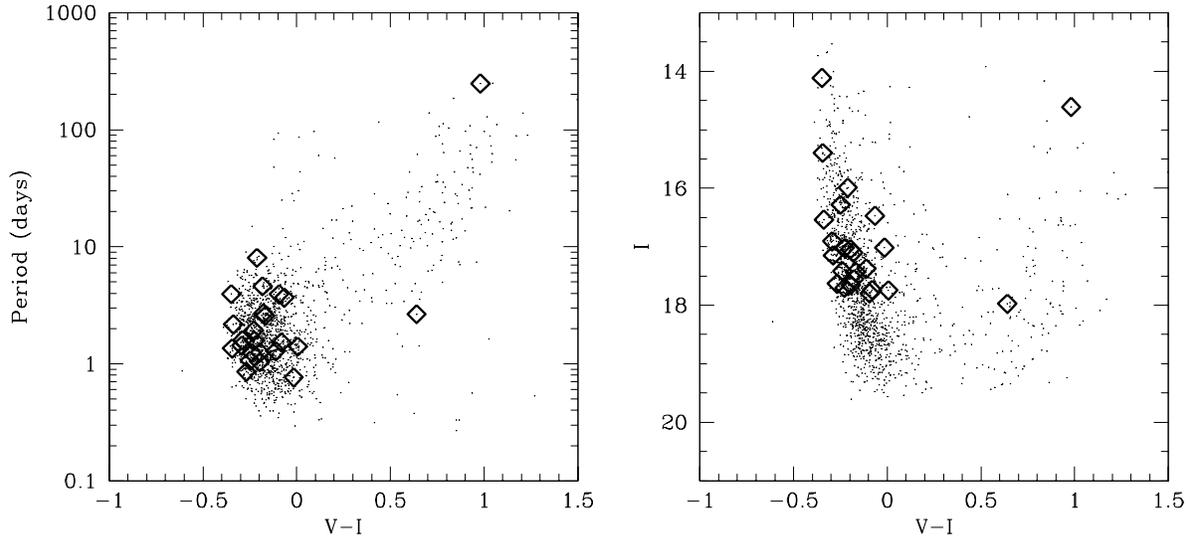}
\caption{\label{ecl_colour_mag}Color - period and color - magnitude
 diagrams for the complete eclipsing binary catalog (small dots, data
 from Udalski et al. 1998). The objects selected to have complete
 eclipses are superimposed (diamonds). The V-I colors have been
 de-reddened assuming E(V-I)=0.14.}
\end{figure}

 Fig.~\ref{ecl_colour_mag} shows color - period and color - magnitude diagrams
 for the selected objects superimposed
 on the full catalog of OGLE SMC eclipsing binaries.
 The V-I colors have been de-reddened assuming E(V-I)=0.14 as before,
 and the systems follow a distribution in color - magnitude space similar to those in the well
 detached catalog. Two of the binaries (smc-sc4 75638 and smc-sc10 137844) are significantly redder
 than the main group. smc-sc10 137844 also appeared in the well detached catalog described in the
 previous sub-section, and the second solution of Tab.~\ref{tab1} is reproduced in Tab.~\ref{tab2}.
 We also obtained a second solution for smc-sc4 75638 assuming more appropriate limb darkening
 coefficients ($x=0.39$, $y=0.20$, from Van Hamme 1993 for an effective temperature of 10000K and a
 $\log g$ of $4.5$). These two solutions are consistent to within standard error. In addition to smc-sc10 137844,
 smc-sc5 129441 also appears in both catalogs. These two binaries are marked by daggers in Tabs.~\ref{tab1}
 and \ref{tab2} and are very special, each having a complete eclipse and small,
 similar components.

\subsubsection{solutions for the Udalski et al. (1998) potential distance indicators} 

\begin{figure}[htbp]
\epsscale{.95}
\plotone{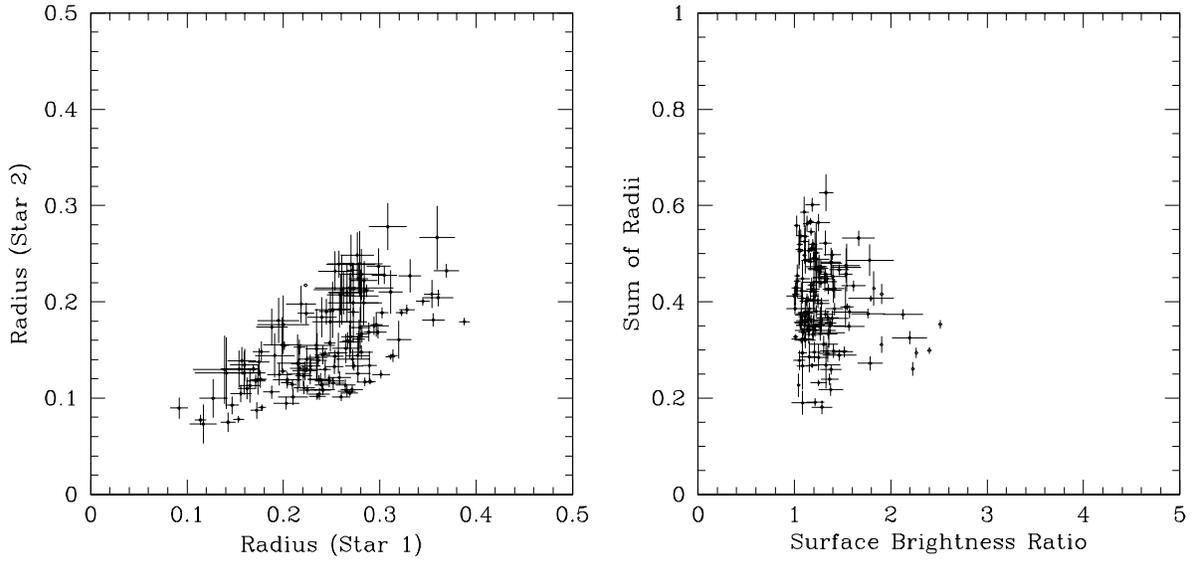}
\caption{\label{OGLE_dists}Solutions for the collection of distance
 indicators complied by Udalski et al. (1998). Left: Solutions for
 the radius of Star 2 plotted against the solutions for the radius
 of Star 1. Only standard errors smaller than 0.05, are shown.
 Right: Solutions for the sum of the radii plotted against
 the solutions for surface brightness ratio. Only standard errors
 in surface brightness ratio smaller than 0.25, and in the sum of
 the radii smaller than 0.1 are plotted.}
\end{figure}

Udalski et al. (1998) selected objects suitable for distance
 determination on the basis of being bright with good photometry and
 having deep and well defined eclipses. For this catalog
Fig.~\ref{OGLE_dists} shows $r_2$ plotted against $r_1$ (left) as well as 
$r_1+r_2$ plotted against surface
 brightness ratio (right). Only standard errors in $r$ smaller than 0.05,
 in $r_1+r_2$ smaller than 0.1, and in surface brightness ratio
 smaller than 0.25 are plotted. Most of these objects have $r's$
greater than 0.25. At that size they will be tidally distorted and so may not
 be ideal as distance indicators since their empirically calibrated surface brightness - temperature
 relations may not be accurate. However, many of the systems with larger components exhibit complete
 eclipses. About half of our samples (marked with
 * in Tabs.~\ref{tab1} and \ref{tab2}) are in the Udalski
 et al. (1998) compilation. Many of these are among the most desirable due
 to their brightness, but the Udalski et al. (1998) catalog is missing
 many of the best systems. This point illustrates the need for quantitative selection.

\section{Conclusion}

We have developed an automated procedure for finding solutions to
 eclipsing binary light curves and applied
 it to the light curves of the 1459 eclipsing binaries found
 in the central 2.4 square degrees of the SMC by the OGLE collaboration
 (Udalski et al. 1998). We find detached solutions for $\sim 80\%$ of these,
 though many systems are found to have large components and may be semi-detached.
 By fitting simulated catalogs we estimate a 95\% success rate in finding acceptable
 converged solutions to detached systems. We extracted two sub-catalogs
 of systems that should make especially good distance indicators. The first is based on
small, well determined fractional radii and surface brightness ratios. The second
 contains systems whose solutions provide accurate radii and indicate complete
 eclipses. Two binaries are in both catalogs and therefore are the leading candidates. One of these comprises
 two G-type, class $II$ giants. 

 The next step is to obtain spectra so as to identify the most suitable systems
on the basis of spectral type. Improved multi-band light curves
should then be observed. In addition, reliable
 distance indicators require  well determined reddening
 (e.g. Gibson 1999; Popowski 2000). Planned observations will address these issues.

\acknowledgements{

 The authors are very greatful to Professor Bohdan Paczynski for suggesting this project, as well
 for his continual encouragement and helpful comments on the manuscript. We would like to thank  
 Professor Slavek Rucinski for comments on a draft version of the manuscript. We would also like to thank
 the referee Dr I. Ribas for a careful reading of our paper. Finally we are indebted to 
 the OGLE collaboration for making their data public domain, and hence projects like this possible.
 This work was supported in part by NSF grants AST-9819787 and AST-9820314 to Professor Paczynski.
 JSBW acknowledges the support of an Australian Postgraduate Award. 
}

\clearpage

\clearpage

\clearpage

\clearpage

\clearpage

\clearpage

\clearpage

\clearpage

\clearpage

\clearpage

\clearpage

\clearpage

\clearpage

\clearpage


\begin{table*}
\begin{center}
\begin{scriptsize}
\begin{tabular}{cc} \hline

 parameter                            &  description \\\hline\hline
 
 $i$        (adjusted*)                &  binary orbit inclination.\\
 $T_1$      (adjusted*)                &  mean surface effective temp. (K) of star 1.\\
 $L_1$      (adjusted*)                &  luminosity for star 1.\\
 $\Omega_1$  (adjusted*)               &  potential of star 1.\\
 $\Omega_2$  (adjusted*)               &  potential of star 2.\\\\
 $T_0$       (adjusted)                &  zero point of orbital ephemeris.\\
 $e$          (adjusted)                &  binary orbit eccentricity.\\\\

 $q$      =  1.0                     &  mass ratio.\\
 $T_2$    =  10000.0                 &  mean surface effective temp. (K) of star 2.\\
 $L_2$      (set from initial guess)    &  luminosity for star 2.\\
 $P$     (from Udalski et al.)     &  period of binary orbit.\\
 $\dot{P}$     =  0.0                     &  first derivative of orbital period.\\
 $\omega_0$=  0.0  or $\pi$           &  initial argument of periastron for star 1. \\
 $\dot{\omega}_0$   =  0.0           &  time derivative of $\omega_0$.     \\                       
 $\lambda_I$=  0.9                   &  wavelength of light curve in microns.\\
 $x_1$      =  0.32                  &  linear limb darkening coefficient of star 1.\\
 $x_2$      =  0.32                  &  linear limb darkening coefficient of star 2.\\
 $y_1$      =  0.18                  &  non-linear limb darkening coefficient of star 1.\\
 $y_2$      =   0.18                 &  non-linear limb darkening coefficient of star 2.\\
 $l_3$      =  0.0                   &  third light.\\
 $f_1$       =  1.0                     &  ratio of axial rotation rate to mean orbital rate.\\
 $f_2$       =  1.0                     &  ratio of axial rotation rate to mean orbital rate.\\
 $g_1$      =  1.0                     &  exponent in gravity brightening (bolo. flux prop. to local gravity).\\
 $g_2$      =  1.0                     &  exponent in gravity brightening (bolo. flux prop. to local gravity).\\
 $A_1$     =  1.000                   &  bolometric albedo of star 1.\\
 $A_2$     =  1.000                   &  bolometric albedo of star 2.\\
 $\lambda$ =  $10^{-5}$              &  the Marquardt multiplier.\\\hline
\end{tabular}
\end{scriptsize}

\caption{\label{tab00} Table of parameters with descriptions. The adjusted parameters are labeled as such
 (those for which convergence is required are also marked by *), and the values of fixed parameters are given.
 Note that g should not be confused with surface gravity.}

\end{center}
\end{table*}

\begin{table*}
\begin{center}
\begin{scriptsize}
\begin{tabular}{cc} \hline

 control integer                    &  description \\\hline\hline
 
 nref     =  1                       &  number of reflections.\\
 mref     =  1                       &  simple reflection treatment.\\
 ld       =  2                       &  logarithmic limb darkening law.\\
 jdphs    =  1                       &  independent variable time.\\
 noise    =  1                       &  scatter scales with sqrt level.\\
 mode     =  2                       &  mode of program operation.\\
 ipb      =  0                       &  for normal operation in mode 2.\\
 ifat1    =  0                       &  for black body (star 1).\\
 ifat2    =  0                       &  for black body (star 2).\\
 n1       =  30                      &  grid size for star 1.\\
 n2       =  30                      &  grid size for star 2.\\
 N1L       =  15                     &  coarse grid integers for star 1.\\
 N2L       =  15                     &  coarse grid integers for star 2.\\
 IFVC1     =  0                      &  no radial velocity curve for star 1. \\
 IFVC2     =  0                      &  no radial velocity curve for star 2.\\
 NLC       =  1                      &  number of light-curves.\\
 KDISK     =  0                      &  no scratch pad.\\
 ISYM      =  1                      &  use symmetrical derivatives.\\\hline

\end{tabular}
\end{scriptsize}

\caption{\label{tab0} Table of control integers with descriptions (nomenclature from Wilson 1998).}

\end{center}
\end{table*}

\begin{table*}
\begin{center}
\begin{scriptsize}
\begin{tabular}{ccc} 

Group 1        &                &                 \\\hline
subset 1       &subset 2       &subset 3       \\\hline\hline
  $e$            &       $i$       &       $T_1$\\
  $\Omega_1$     &   $L_1$         &        $T_o$   \\
  $\Omega_2$     &                 &                \\\hline\\\\
Group 2        &                &                 \\\hline
subset 1       &subset 2       &subset 3       \\\hline\hline
  $e$            &       $i$       &       $T_1$\\
  $T_o$     &   $\Omega_2$         &        $\Omega_1$   \\
  $L_1$     &                 &        \\\hline

\end{tabular}
\end{scriptsize}

\caption{\label{tab01} Table showing the two groups of subsets used by dc.}

\end{center}
\end{table*}


\begin{table*}
\begin{center}
\begin{tiny}
\begin{tabular}{ccccccccccc} \hline
        &        & Radius of              & Radius of               & Surf.-bright.            & inclination & eccentricity &Period  &       &     &\\
Field   &  Object& star 1 $\frac{R_1}{a}$ & star 2 $\frac{R_2}{a}$  &  ratio $\frac{J_1}{J_2}$ &$i$ (degrees)& $e$          &(days)  &  I    & B-V & V-I  \\\hline\hline
3 & 157431 & 0.133 $\pm$ 0.032 &  0.109 $\pm$ 0.042 &  1.435 $\pm$ 0.248  &  84.3 $\pm$ 1.7  &  0.005 $\pm$ 0.003  & 2.37674 & 18.071 & -0.042 & -0.031\\
4 & 19130 & 0.153 $\pm$ 0.021 &  0.097 $\pm$ 0.035 &  1.263 $\pm$ 0.222  &  84.7 $\pm$ 2.3  &  0.007 $\pm$ 0.004  & 5.95288 & 18.280 & 0.002 & 0.058\\
4 & 23166 & 0.095 $\pm$ 0.032 &  0.088 $\pm$ 0.031 &  1.226 $\pm$ 0.180  &  89.6 $\pm$ 6.2  &  0.002 $\pm$ 0.002  & 4.04442 & 18.132 & -0.109 & -0.048\\
4 & 26278* & 0.199 $\pm$ 0.020 &  0.155 $\pm$ 0.029 &  1.268 $\pm$ 0.085  &  86.4 $\pm$ 2.3  &  0.007 $\pm$ 0.003  & 3.06268 & 17.343 & -0.029 & 0.015\\
4 & 54029* & 0.154 $\pm$ 0.020 &  0.130 $\pm$ 0.019 &  1.253 $\pm$ 0.100  &  85.0 $\pm$ 0.9  &  0.104 $\pm$ 0.002  & 2.61830 & 17.815 & -0.038 & -0.020\\
5 & 129441*$\dagger$ & 0.199 $\pm$ 0.004 &  0.128 $\pm$ 0.003 &  1.012 $\pm$ 0.026  &  87.3 $\pm$ 0.8  &  0.375 $\pm$ 0.001  & 8.05053 & 15.991 & -0.128 & -0.071\\
5 & 180516* & 0.114 $\pm$ 0.006 &  0.077 $\pm$ 0.005 &  1.216 $\pm$ 0.093  &  88.4 $\pm$ 1.0  &  0.037 $\pm$ 0.002  & 7.58354 & 17.491 & -0.099 & -0.089\\
5 & 191183 & 0.159 $\pm$ 0.022 &  0.126 $\pm$ 0.023 &  1.152 $\pm$ 0.096  &  84.8 $\pm$ 1.1  &  0.178 $\pm$ 0.003  & 3.28794 & 18.212 & -0.024 & 0.012\\
6 & 11143* & 0.147 $\pm$ 0.007 &  0.093 $\pm$ 0.009 &  1.366 $\pm$ 0.093  &  85.9 $\pm$ 0.9  &  0.161 $\pm$ 0.002  & 5.72586 & 16.164 & -0.129 & -0.101\\
6 & 67289* & 0.157 $\pm$ 0.015 &  0.139 $\pm$ 0.014 &  1.207 $\pm$ 0.058  &  86.6 $\pm$ 0.6 &  0.068 $\pm$ 0.002  & 3.29158 & 15.939 & -0.231 & -0.035\\
6 & 191016 & 0.193 $\pm$ 0.020 &  0.134 $\pm$ 0.015 &  1.108 $\pm$ 0.132  &  87.2 $\pm$ 3.2  &  0.156 $\pm$ 0.006  & 1.92518 & 18.827 & -0.018 & 0.158\\
6 & 222388 & 0.131 $\pm$ 0.013 &  0.077 $\pm$ 0.009 &  1.098 $\pm$ 0.147  &  89.1 $\pm$ 5.3  &  0.014 $\pm$ 0.004   & 6.61099 & 18.627 & 0.027 & 0.162\\
7 & 13487* & 0.195 $\pm$ 0.022 &  0.181 $\pm$ 0.024 &  1.068 $\pm$ 0.052  &  83.8 $\pm$ 0.5  &  0.021 $\pm$ 0.002  & 5.71080 & 15.709 & -0.036 & -0.113\\
7 & 83110* & 0.165 $\pm$ 0.010 &  0.113 $\pm$ 0.018 &  1.051 $\pm$ 0.062  &  86.2 $\pm$ 1.7  &  0.225 $\pm$ 0.002  & 5.29831 & 17.238 & -0.190 & -0.149\\
8 & 35399 & 0.135 $\pm$ 0.025 &  0.095 $\pm$ 0.042 &  1.200 $\pm$ 0.184  &  86.2 $\pm$ 3.5  &  0.315 $\pm$ 0.003  & 4.68531 & 18.282 & -0.050 & -0.008\\
8 & 39206* & 0.196 $\pm$ 0.013 &  0.125 $\pm$ 0.012 &  1.110 $\pm$ 0.109  &  86.8 $\pm$ 2.7  &  0.125 $\pm$ 0.004  & 3.92190 & 18.042 & -0.005 & 0.074\\
8 & 53788* & 0.141 $\pm$ 0.033 &  0.126 $\pm$ 0.037 &  1.086 $\pm$ 0.065  &  87.6 $\pm$ 1.7  &  0.041 $\pm$ 0.003  & 5.10280 & 18.108 & -0.090 & -0.013\\
9 & 55949 & 0.155 $\pm$ 0.009 &  0.105 $\pm$ 0.009 &  0.946 $\pm$ 0.074  &  88.0 $\pm$ 2.2  &  0.005 $\pm$ 0.002  & 63.98542 & 17.459 & 0.834 & 0.926\\
  &       & (0.156 $\pm$ 0.008 &  0.106 $\pm$ 0.012 &  0.924 $\pm$ 0.078  &  87.5 $\pm$ 2.5 &  0.004 $\pm$ 0.002)&         &        &        &       \\ 
9 & 67706* & 0.173 $\pm$ 0.008 &  0.119 $\pm$ 0.016 &  1.345 $\pm$ 0.088  &  85.3 $\pm$ 1.6  &  0.163 $\pm$ 0.002  & 3.25213 & 16.912 & -0.157 & -0.121\\
9 & 89664* & 0.139 $\pm$ 0.033 &  0.130 $\pm$ 0.036 &  1.327 $\pm$ 0.149  &  86.8 $\pm$ 1.7  &  0.021 $\pm$ 0.003  & 2.34910 & 18.088 & -0.157 & -0.092\\
10 & 22190* & 0.191 $\pm$ 0.020 &  0.144 $\pm$ 0.023 &  1.358 $\pm$ 0.163  &  83.7 $\pm$ 1.6  &  0.040 $\pm$ 0.003  & 1.44831 & 18.025 & -0.120 & -0.049\\
10 & 39758* & 0.175 $\pm$ 0.011 &  0.138 $\pm$ 0.014 &  1.303 $\pm$ 0.148  &  89.0 $\pm$ 4.9  &  0.019 $\pm$ 0.003  & 3.35275 & 15.882 & -0.229 & -0.155\\
10 & 68458* & 0.160 $\pm$ 0.016 &  0.135 $\pm$ 0.017 &  1.082 $\pm$ 0.046  &  85.6 $\pm$ 0.5  &  0.049 $\pm$ 0.002  & 3.40799 & 17.284 & -0.190 & -0.149\\
10 & 108195* & 0.188 $\pm$ 0.016 &  0.174 $\pm$ 0.019 &  1.278 $\pm$ 0.080  &  81.9 $\pm$ 0.5  &  0.001 $\pm$ 0.003  & 2.26354 & 17.206 & -0.150 & -0.110\\
10 & 137844$\dagger$ & 0.186 $\pm$ 0.003 &  0.147 $\pm$ 0.003 &  1.103 $\pm$ 0.038  &  89.3 $\pm$ 1.5  &  0.008 $\pm$ 0.001  & 248.00000 & 14.609 & 1.016 & 1.121\\
   &        & (0.192 $\pm$ 0.004 &  0.148 $\pm$ 0.004 &  1.022 $\pm$ 0.029   & 88.7 $\pm$ 1.0 & 0.008$\pm$0.001)    &           &        &       &      \\\hline

\end{tabular}
\end{tiny}

\caption{\label{tab1} Table of parameters for eclipsing binaries
 selected as likely distance indicators on the basis of being well detached. The periods,
 I-magnitudes and colors are taken from Udalski et al. (1998). Those marked with an asterisk
 appeared in the list complied by Udalski et al. (1998). The solutions marked with a dagger also 
 appear in Tab.~\ref{tab2}. Second solutions
 for the two red, long period objects assuming different limb-darkening
 coefficients are given in parentheses.}

\end{center}
\end{table*}


\begin{table*}
\begin{center}
\begin{tiny}
\begin{tabular}{ccccccccccc} \hline
        &        & Radius of              & Radius of               & Surf.-bright.            & inclination & eccentricity &Period  &       &     &\\
Field   &  Object& star 1 $\frac{R_1}{a}$ & star 2 $\frac{R_2}{a}$  &  ratio $\frac{J_1}{J_2}$ &$i$ (degrees)& $e$          &(days)  &  I    & B-V & V-I  \\\hline\hline

3 & 193413 & 0.400 $\pm$ 0.012 &  0.192 $\pm$ 0.005 &  1.392 $\pm$ 0.087  &  84.3 $\pm$ 1.7  &  0.011 $\pm$ 0.002  & 0.76458 & 17.013 & -0.001 & 0.124 \\
4 & 75638* & 0.281 $\pm$ 0.012 &  0.148 $\pm$ 0.007 &  1.022 $\pm$ 0.056  &  84.7 $\pm$ 1.8  &  0.068 $\pm$ 0.003  & 2.65340 & 17.972 & 0.162 & 0.780 \\
  &       & (0.277 $\pm$ 0.012&   0.145 $\pm$ 0.012&  1.016 $\pm$ 0.060   &  85.2$\pm$1.8   &   0.068 $\pm$ 0.003)&          &       &       &        \\  
4 & 103706* & 0.388 $\pm$ 0.006 &  0.179 $\pm$ 0.004 &  1.166 $\pm$ 0.035  &  82.4 $\pm$ 0.9  &  0.006 $\pm$ 0.002  & 1.35588 & 15.395 & -0.211 & -0.205 \\
4 & 175333* & 0.253 $\pm$ 0.009 &  0.143 $\pm$ 0.007 &  1.278 $\pm$ 0.084  &  85.0 $\pm$ 1.5  &  0.012 $\pm$ 0.003  & 1.25116 & 17.653 & -0.125 & -0.056 \\
5 & 16658* & 0.345 $\pm$ 0.007 &  0.200 $\pm$ 0.004 &  1.174 $\pm$ 0.040  &  85.7 $\pm$ 1.3  &  0.002 $\pm$ 0.002  & 1.24619 & 17.414 & -0.148 & -0.100 \\
5 & 43860 & 0.422 $\pm$ 0.016 &  0.173 $\pm$ 0.009 &  1.069 $\pm$ 0.087  &  81.9 $\pm$ 2.0  &  0.005 $\pm$ 0.005  & 1.40380 & 17.745 & 0.085 & 0.146 \\
5 & 129441*$\dagger$ & 0.199 $\pm$ 0.004 &  0.128 $\pm$ 0.003 &  1.012 $\pm$ 0.026  &  87.3 $\pm$ 0.8  &  0.375 $\pm$ 0.001  & 8.05053 & 15.991 & -0.128 & -0.071 \\
5 & 196565* & 0.204 $\pm$ 0.006 &  0.116 $\pm$ 0.003 &  1.074 $\pm$ 0.036  &  88.7 $\pm$ 2.5  &  0.085 $\pm$ 0.002  & 3.94260 & 16.910 & 9.999 & 9.999 \\
5 & 266131 & 0.332 $\pm$ 0.008 &  0.216 $\pm$ 0.009 &  1.363 $\pm$ 0.067  &  85.1 $\pm$ 1.5  &  0.003 $\pm$ 0.003  & 1.30288 & 17.017 & -0.051 & -0.014 \\
5 & 283079* & 0.272 $\pm$ 0.006 &  0.233 $\pm$ 0.007 &  1.055 $\pm$ 0.027  &  88.5 $\pm$ 1.8  &  0.001 $\pm$ 0.001  & 1.28358 & 17.374 & -0.074 & 0.028 \\
5 & 300727* & 0.298 $\pm$ 0.009 &  0.169 $\pm$ 0.006 &  1.465 $\pm$ 0.100  &  87.0 $\pm$ 2.6  &  0.003 $\pm$ 0.003  & 1.57142 & 17.168 & -0.060 & 0.025 \\
6 & 77359* & 0.228 $\pm$ 0.010 &  0.133 $\pm$ 0.007 &  1.170 $\pm$ 0.079  &  85.5 $\pm$ 1.4  &  0.101 $\pm$ 0.003  & 3.68834 & 16.472 & -0.175 & 0.074 \\
6 & 158178 & 0.207 $\pm$ 0.006 &  0.123 $\pm$ 0.004 &  1.204 $\pm$ 0.062  &  88.1 $\pm$ 2.1  &  0.242 $\pm$ 0.002  & 2.16926 & 16.538 & -0.201 & -0.199 \\
6 & 180306* & 0.201 $\pm$ 0.004 &  0.155 $\pm$ 0.004 &  1.374 $\pm$ 0.061  &  88.3 $\pm$ 1.0  &  0.042 $\pm$ 0.002  & 4.59712 & 17.099 & -0.087 & -0.042 \\
6 & 215965 & 0.391 $\pm$ 0.006 &  0.171 $\pm$ 0.004 &  1.099 $\pm$ 0.034  &  83.0 $\pm$ 1.0  &  0.004 $\pm$ 0.002  & 3.94604 & 14.116 & -0.204 & -0.209 \\
6 & 242137 & 0.373 $\pm$ 0.008 &  0.179 $\pm$ 0.005 &  1.187 $\pm$ 0.049  &  82.4 $\pm$ 1.0  &  0.008 $\pm$ 0.003  & 2.45717 & 17.204 & 9.999 & 9.999 \\
6 & 319966 & 0.392 $\pm$ 0.009 &  0.170 $\pm$ 0.006 &  1.098 $\pm$ 0.050  &  81.1 $\pm$ 1.2  &  0.012 $\pm$ 0.003  & 1.06463 & 16.282 & -0.142 & -0.109 \\
7 & 115361* & 0.328 $\pm$ 0.008 &  0.192 $\pm$ 0.005 &  1.192 $\pm$ 0.048  &  87.6 $\pm$ 2.8  &  0.016 $\pm$ 0.002  & 1.57361 & 17.018 & -0.106 & -0.086 \\
7 & 133897* & 0.273 $\pm$ 0.006 &  0.190 $\pm$ 0.004 &  1.266 $\pm$ 0.040  &  88.4 $\pm$ 2.0  &  0.028 $\pm$ 0.002  & 1.59029 & 17.147 & -0.162 & -0.149 \\
7 & 255621* & 0.215 $\pm$ 0.007 &  0.136 $\pm$ 0.005 &  1.255 $\pm$ 0.059  &  88.9 $\pm$ 2.8  &  0.144 $\pm$ 0.002  & 4.33063 & 16.180 & -0.177 & -0.111 \\
7 & 255917* & 0.356 $\pm$ 0.012 &  0.181 $\pm$ 0.007 &  1.067 $\pm$ 0.052  &  82.9 $\pm$ 1.4  &  0.004 $\pm$ 0.004  & 1.52412 & 17.737 & -0.022 & 0.059 \\
8 & 42722 & 0.206 $\pm$ 0.008 &  0.121 $\pm$ 0.007 &  1.066 $\pm$ 0.079  &  86.4 $\pm$ 1.5  &  0.058 $\pm$ 0.002  & 2.69232 & 17.551 & -0.091 & -0.037 \\
8 & 76421 & 0.316 $\pm$ 0.008 &  0.179 $\pm$ 0.010 &  1.440 $\pm$ 0.110  &  83.5 $\pm$ 1.6  &  0.003 $\pm$ 0.003  & 1.92885 & 17.703 & -0.128 & -0.088 \\
8 & 119647 & 0.433 $\pm$ 0.016 &  0.182 $\pm$ 0.006 &  1.065 $\pm$ 0.059  &  84.3 $\pm$ 2.5  &  0.004 $\pm$ 0.004  & 0.85494 & 17.631 & -0.092 & -0.129 \\
8 & 163828 & 0.223 $\pm$ 0.013 &  0.128 $\pm$ 0.010 &  1.107 $\pm$ 0.093  &  85.2 $\pm$ 1.8  &  0.207 $\pm$ 0.004  & 3.93281 & 17.794 & -0.076 & 0.046 \\
10 & 114 & 0.398 $\pm$ 0.012 &  0.164 $\pm$ 0.006 &  1.105 $\pm$ 0.072  &  81.3 $\pm$ 1.5  &  0.001 $\pm$ 0.004  & 1.03348 & 17.041 & -0.138 & -0.057 \\
10 & 6166 & 0.302 $\pm$ 0.007 &  0.189 $\pm$ 0.006 &  1.202 $\pm$ 0.046  &  84.8 $\pm$ 1.1  &  0.014 $\pm$ 0.001  & 2.57091 & 17.431 & -0.078 & -0.029 \\
10 & 78731* & 0.322 $\pm$ 0.006 &  0.189 $\pm$ 0.004 &  1.178 $\pm$ 0.037  &  86.8 $\pm$ 1.5  &  0.013 $\pm$ 0.002  & 1.42945 & 16.904 & -0.196 & -0.154 \\
10 & 137844$\dagger$ & 0.186 $\pm$ 0.003 &  0.147 $\pm$ 0.003 &  1.103 $\pm$ 0.038  &  89.3 $\pm$ 1.5  &  0.008 $\pm$ 0.001  & 248.00000 & 14.609 & 1.016 & 1.121 \\
   &        & (0.192 $\pm$ 0.004 &  0.148 $\pm$ 0.004 &  1.022 $\pm$ 0.029   &  88.7$\pm$1.0  & 0.008$\pm$0.001)    &           &        &       &       \\\hline

\end{tabular}
\end{tiny}
\caption{\label{tab2} Table of parameters for eclipsing binaries
 selected as likely distance indicators based on their having complete eclipses.
 The periods, I-magnitudes and
 colors are taken from Udalski et al. (1998). Those marked with an asterisk
 appeared in the list complied by Udalski et al. (1998). The solutions marked with a dagger also 
 appear in Tab.~\ref{tab1}. Second solutions
 for the red objects 4-75638 and 10-137844 assuming different limb-darkening
 coefficients are given in parentheses.}

\end{center}
\end{table*}

\end{document}